%% file: main.tex
\documentclass[sigconf]{acmart}

\usepackage{amsmath}
\usepackage{booktabs}
\usepackage{graphicx}
\usepackage{multirow}
\usepackage{tabularx}
\usepackage{placeins}
\usepackage{enumitem}
\usepackage{xcolor}
\usepackage{microtype}
\microtypesetup{expansion=false}
\makeatletter
\AtBeginMaketitle{%
  \let\@authornotes\@empty
  \pretocmd{\@copyrightpermission}{\textsuperscript{*}Corresponding author.\par}{}{}%
}
\makeatother
\renewcommand{\textbullet}{\textendash}
\copyrightyear{2027}
\acmYear{2027}
\setcopyright{cc}
\setcctype{by}
\acmConference[WSDM '27]{Proceedings of the Twentieth ACM International Conference on Web Search and Data Mining}{February 15--19, 2027}{Hong Kong, Hong Kong}
\acmBooktitle{Proceedings of the Twentieth ACM International Conference on Web Search and Data Mining (WSDM '27), February 15--19, 2027, Hong Kong, Hong Kong}
\acmDOI{10.1145/3862322.3863599}
\acmISBN{979-8-4007-3117-4/2027/02}

\begin{document}

\title{Grounding User Preferences in Shared Multimodal Concepts for Recommendation}
\input{author_info}
\renewcommand{\shortauthors}{Ao Xu et al.}

\begin{abstract}
Multimodal recommendation has substantially advanced item understanding by incorporating rich visual and textual content. Yet user preferences are still commonly encoded as implicit signals tied to individual items, leaving recurring semantic factors without persistent identities that can be shared across items. We introduce MKGE, a multimodal knowledge graph framework that represents these factors as canonical, addressable concepts. MKGE induces a domain-adaptive concept vocabulary from item images and titles, grounds each user's interaction history in supported concepts, and jointly models direct user--item evidence and concept-mediated preference signals. By turning recurring semantics into shared coordinates, MKGE enables evidence from different items to accumulate around the same preference factors. Extensive experiments on three public benchmarks demonstrate that MKGE consistently outperforms collaborative, knowledge-graph, multimodal, and recent LLM/MLLM-based recommenders across all six reported metrics. Controlled studies further demonstrate the importance of canonical concept identities and personalized user--concept correspondence, with especially pronounced gains over removing or rewiring the user graph for low-activity users. These results establish canonical multimodal concepts as effective units for preference modeling and transfer. Code, constructed knowledge graphs, and prompts are available at \url{https://github.com/Garden1a-X/MKGE}.
\end{abstract}

\ccsdesc[500]{Information systems~Recommender systems}
\ccsdesc[300]{Computing methodologies~Knowledge representation and reasoning}

\keywords{multimodal recommendation, multimodal large language models, knowledge graphs, preference grounding}

\maketitle

\section{Introduction}

Multimodal recommendation has substantially advanced how recommender systems represent items. Images and text reveal appearance, style, atmosphere, and theme beyond what can be learned from interaction IDs alone. Successive advances in multimodal feature learning, graph construction, and self-supervised learning have made these signals increasingly effective for recommendation~\cite{He_McAuley_2016,10.1145/3343031.3351034,10.1145/3474085.3475259,10.1145/3543507.3583251,10.1145/3581783.3611943}. As a result, recommenders can now describe the content of an item at an unprecedented semantic resolution.

This progress, however, exposes a growing asymmetry in user--item matching: item representations have become increasingly fine-grained, while user preference is still largely inferred at the granularity of whole-item interactions. A click or purchase indicates which item a user chose, yet leaves the factors that motivated the choice entangled with all other properties of that item. User-aware weighting and target-conditioned attention can emphasize relevant parts of multimodal representations~\cite{kim2024monet,wen2024unified}, but the resulting preference factors remain implicit within continuous representations, without persistent identities that can be directly addressed and shared across items.

\begin{figure}[t]
  \centering
  \includegraphics[width=.98\columnwidth]{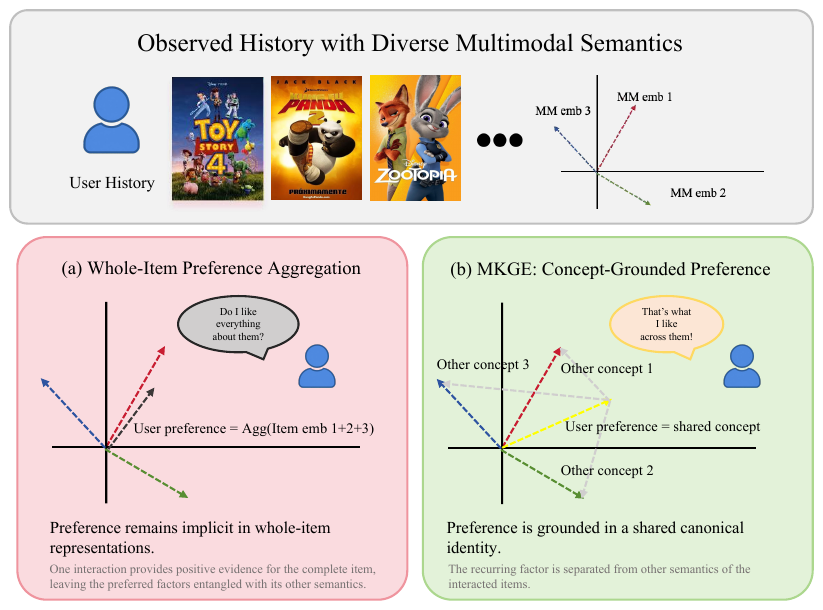}
  \caption{Why user preference should be grounded beyond whole items. A user's history contains items with diverse multimodal semantics. (a) Whole-item aggregation leaves a recurring preferred factor implicit and entangled with other item semantics. (b) MKGE gives that factor a shared canonical identity, making it an explicit coordinate for user preference.}
  \Description{A motivation diagram with an observed history above and two preference views below. Whole-item preference aggregation combines the embeddings of diverse interacted items, leaving the preferred factor implicit. MKGE instead grounds the user in a shared canonical concept that recurs across the items.}
  \label{fig:motivation}
\end{figure}

Consider a user whose history contains diverse items that nevertheless share a recurring multimodal factor. Treating each interaction as positive evidence for the whole item leaves that preferred factor entangled with the item's other semantics, as illustrated in Figure~\ref{fig:motivation}(a). Giving the recurring factor a canonical identity instead provides a shared coordinate that persists across otherwise different items, as shown in Figure~\ref{fig:motivation}(b). The user's preference can then be grounded in what recurs across the history rather than represented only through an aggregation of whole-item embeddings. This shared identity changes the unit of user modeling from whole items alone to the multimodal factors repeatedly supported by their content.

A shared-coordinate representation raises three challenges. First, its semantic units must cover domain-specific and compositional multimodal factors while resolving equivalent observations to consistent identities across items. Second, shared item semantics must be converted into personalized preference: among all concepts present in a user's interacted items, the model must identify those supported by the history at different temporal scales. Third, concept-mediated evidence must work together with direct collaborative signals, which remain indispensable for capturing behavioral regularities beyond observable content.

Multimodal large language models provide the capacity to recognize open-ended, domain-specific factors from item images and titles, while knowledge graphs provide persistent identities through which these factors can be shared. Building on these complementary strengths, we propose \emph{MKGE}, a multimodal knowledge graph framework for grounded user preference learning. MKGE induces a domain-adaptive vocabulary of canonical concepts, grounds each user's chronological history in the same vocabulary, and learns a dual-path recommender over direct user--item interactions and concept-mediated user--concept--item relations. In this structure, the item graph establishes reusable semantic coordinates and the user graph records which coordinates are supported by each user's interaction history.

\begingroup\sloppy
This formulation leads to concrete empirical questions. Does canonical identity strengthen the reuse of recurring semantics? Does recommendation depend on the correspondence between each user and their grounded concepts, beyond concept frequency, global distinctiveness, or graph connectivity? Do concept-mediated paths complement direct collaborative evidence, and are they especially useful when a user provides fewer interactions? Across three benchmarks, MKGE leads collaborative, knowledge-aware, multimodal, and recent large-model baselines on all six metrics. Controlled studies support shared concept identity and personalized grounding, with the largest gains over removing or rewiring the user graph for low-activity users.
\par\endgroup

In summary, we make the following contributions:
\begin{itemize}[leftmargin=*,nosep]
  \item To establish shared semantic identities, we formulate \emph{grounded multimodal preference learning} and induce a domain-adaptive vocabulary in which recurring multimodal factors become canonical, addressable coordinates across items.
  \item To turn shared item semantics into user preference, we introduce temporal history grounding and a dual-path recommender that combines short- and long-term user--concept relations with direct collaborative evidence.
  \item To determine when and why these representations improve recommendation, we conduct controlled grounding, activity, component, canonicalization, case, robustness, and transfer studies across three benchmarks. The results verify the importance of preserving user--concept correspondence, shared concept identity, and complementary recommendation paths.
\end{itemize}

\section{Related Work}

\subsection{Multimodal Representation Learning for Recommendation}
\begingroup\sloppy
Multimodal recommendation enriches interaction signals with item content. VBPR projects visual features into a latent space and learns user-specific visual factors for personalized ranking~\cite{He_McAuley_2016}. MMGCN subsequently propagates modality-specific information over user--item graphs to learn both user and item representations~\cite{10.1145/3343031.3351034}. These models establish the central value of content-aware matching: behavior determines which directions in a multimodal representation space are useful for recommendation.

Later work improves the structure, supervision, and reliability of this representation process. LATTICE learns modality-aware item--item graphs from content~\cite{10.1145/3474085.3475259}; BM3 bootstraps latent views and aligns modalities without auxiliary graph augmentation or sampled negatives~\cite{10.1145/3543507.3583251}; and FREEDOM freezes the multimodal item graph while denoising user--item interactions~\cite{10.1145/3581783.3611943}. SMORE suppresses modality-specific and cross-modal noise during fusion~\cite{ong2025smore}, and GUIDER recalibrates noisy feedback and distills modality representations toward recommendation semantics~\cite{li2025guider}. These approaches learn continuous user and item representations by improving graph structure, supervision, and modality fusion.
\endgroup

\subsection{Multimodal User Preference Modeling}
\begingroup\sloppy
Several multimodal recommenders focus more directly on the user side of the match. MONET strengthens the contribution of modality features to final embeddings and uses target-aware attention to construct a user representation conditioned on the candidate item~\cite{kim2024monet}. Tavern similarly derives a visual preference signal for each historical click from its shared visual information with the target, removes irrelevant information through selective disentanglement, and aggregates the resulting signals with a recurrent encoder~\cite{wen2024unified}. Both demonstrate that a user's relevant multimodal evidence depends on which parts of the history are activated for the current decision.

Large models provide another interface for preference modeling. TALLRec establishes instruction tuning as an efficient way to align an LLM with recommendation data~\cite{Bao2023TALLRec}. MLLM-MSR converts item images into textual summaries, recurrently summarizes the evolution of user preference, and fine-tunes an MLLM for sequential recommendation~\cite{Ye_Zheng_Shen_Wang_Zhang_Zhu_Yu_Zhang_Xiong_2025}. MLLMRec converts images into semantic descriptions and reasons over each behavioral description list to obtain a purified user preference profile, while also refining the item--item graph~\cite{dang2026mllmrecpreferencereasoningparadigm}. These methods move beyond uniform history pooling through target-conditioned representations or language-level user profiles. Recent work further develops explicit preference profiles through KG-augmented conversational reasoning~\cite{qiu2025compass}, multimodal multi-interest profiling~\cite{wang2026prime}, and online natural-language interest personas~\cite{wang2026personas}.

MKGE represents preference through a persistent factor-level interface. A concept discovered in one item retains the same identity when it recurs elsewhere. A history selects supported concepts, and a candidate connects to the user through those same nodes, allowing preference evidence to accumulate across otherwise different items.
\endgroup

\subsection{Shared Semantic Structures for Recommendation}
Knowledge-aware recommenders make side information reusable through entities, relations, and paths. RippleNet expands a user's clicked items along KG links to form candidate-aware preference distributions~\cite{10.1145/3269206.3271739}. KGAT propagates attentive messages over a unified collaborative knowledge graph~\cite{10.1145/3292500.3330989}, while KGIN represents latent user intents as attentive combinations of KG relations and preserves relational paths during propagation~\cite{10.1145/3442381.3450133}. These methods show how shared entities can transfer preference evidence beyond direct item co-occurrence.

Multimodal and large-model methods enrich the structures available for such propagation. MKGAT encodes visual and textual information for entities and aggregates their neighbors in a multimodal KG~\cite{10.1145/3340531.3411947}. DOGE uses LLM-enhanced cross-modal features to construct a hyper-knowledge graph containing user--user, item--item, and higher-order relations~\cite{Meng_Meng_Jin_Lin_Wu_2025}. MMQ quantizes multimodal item content into semantic IDs and behaviorally adapts those codes for recommendation~\cite{xu2026mmq}. These methods make multimodal semantics accessible to graph propagation or discrete sequential modeling.

LLM-based information extraction and KG construction provide relevant precedents for creating semantic structure. Text2DB adapts extraction to a target database schema~\cite{jiao-etal-2024-text2db}; SAC-KG uses generation, verification, and pruning to construct domain KGs~\cite{chen-etal-2024-sac}; and IncPrompt incrementally induces and verifies open-domain event schemas~\cite{li-etal-2023-open}. MKGE specializes semantic induction to multimodal recommendation by discovering domain-relevant factors, normalizing them into canonical concept nodes, connecting the catalog to those nodes, and grounding user histories in the same coordinate system.

\section{Method}
\label{sec:method}

\subsection{Problem Formulation and Overview}

Let $\mathcal{U}$, $\mathcal{I}$, $\mathcal{E}$, and $\mathcal{T}$ denote users, items, canonical multimodal concepts, and semantic relation types. A multimodal concept is a reusable semantic factor inferred jointly from an item's image and short title; examples include style, atmosphere, lighting, product appearance, and thematic cues. Each item $i$ has an image--title input $x_i$. Chronological implicit feedback is $\mathbf{R}\in\{0,1\}^{|\mathcal{U}|\times|\mathcal{I}|}$, and $\mathbf{R}^{t}$ restricts it to interactions in temporal bucket $t$. For relation $r$, induction produces $\mathbf{A}^{r}\in\{0,1\}^{|\mathcal{I}|\times|\mathcal{E}|}$, where $A^{r}_{ie}=1$ iff concept $e$ describes item $i$ under relation $r$. We write $\mathbf{A}=\bigvee_{r\in\mathcal{T}}\mathbf{A}^{r}$. Grounding produces short- and long-term matrices $\mathbf{P}^{S},\mathbf{P}^{L}\in\{0,1\}^{|\mathcal{U}|\times|\mathcal{E}|}$.

The induced vocabulary provides an explicit semantic matching operator. For horizon $q\in\{S,L\}$,
\begin{equation}
\mathbf{M}^{q}=\mathbf{P}^{q}\mathbf{D}_{E}^{-1}\mathbf{A}^{\top},
\qquad M^{q}_{ui}=\sum_{e}\frac{P^{q}_{ue}A_{ie}}{d_e},
\label{eq:semantic-match}
\end{equation}
where $d_e=\sum_i A_{ie}$ and $\mathbf{D}_E=\operatorname{diag}(d_e)$. Thus $M^{q}_{ui}$ is a degree-normalized count of grounded concepts shared by user $u$ and candidate $i$. Eq.~\eqref{eq:semantic-match} makes shared-concept matching explicit. The recommender learns relation-specific propagation over these paths and combines the resulting representations with collaborative signals. Canonicalization keeps repeated factors in the same column and preserves their path contribution.

\begin{figure*}[t]
  \centering
  \includegraphics[width=.84\textwidth]{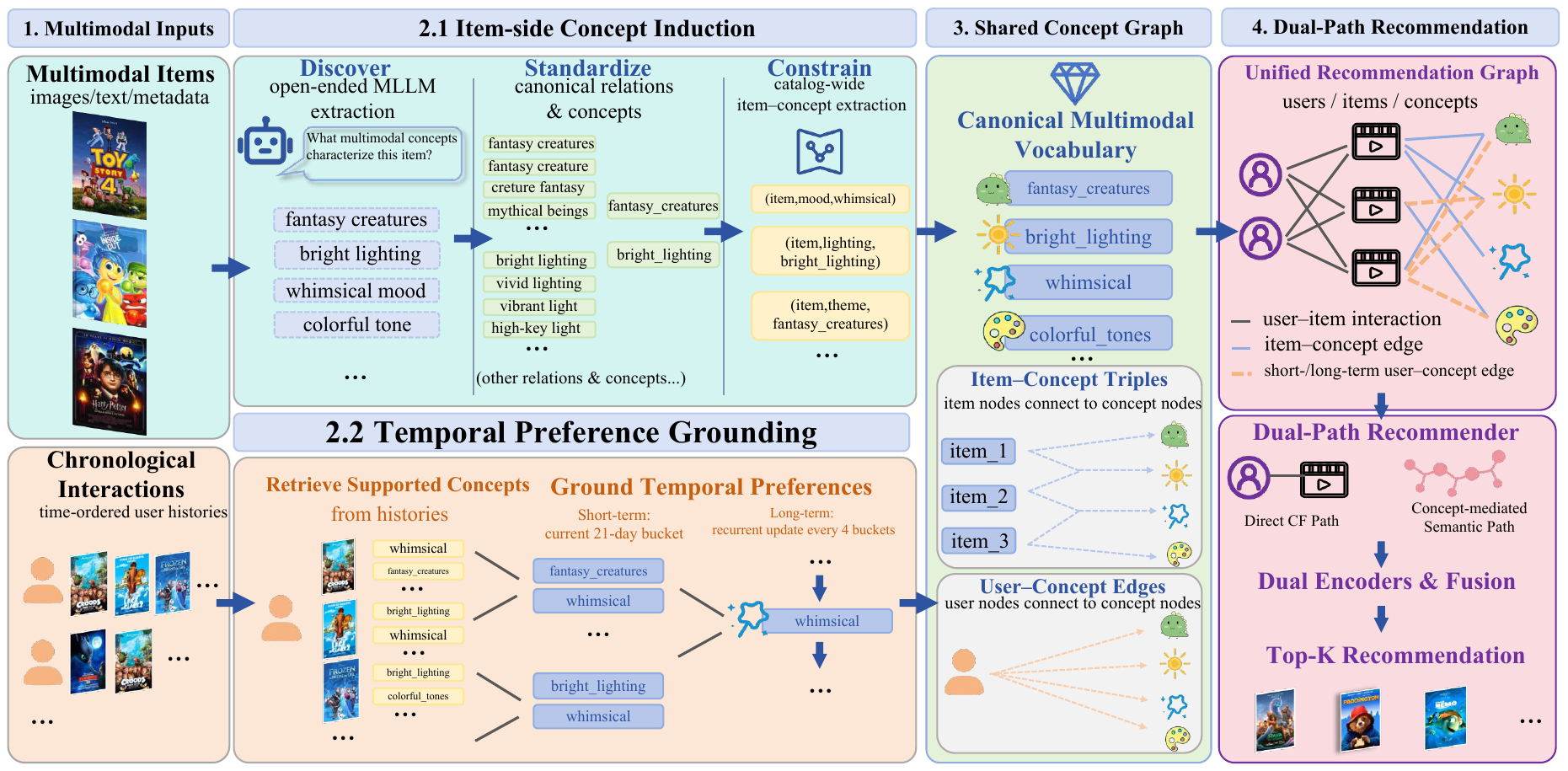}
  \caption{Overview of MKGE. Discover--Standardize--Constrain induces a canonical multimodal vocabulary and catalog-wide item--concept triples. Temporal preference grounding links short- and long-term user interests to the same concepts, forming a unified recommendation graph for direct collaborative and concept-mediated semantic paths.}
  \Description{The framework takes multimodal item content and time-ordered user histories as inputs. Discover, Standardize, and Constrain induce a canonical vocabulary and catalog-wide item--concept triples. Temporal grounding links short- and long-term user preferences to supported concepts. Item--concept and user--concept edges then form a unified graph whose direct collaborative and concept-mediated semantic paths are fused for recommendation.}
  \label{fig:framework}
\end{figure*}

Fig.~\ref{fig:framework} organizes MKGE around this information flow. Concept induction creates the shared coordinate system, preference grounding connects behavior to it, and the recommender learns direct and semantic paths. MLLM operations occur offline and can be parallelized; online ranking uses only the trained graph model.

\subsection{Domain-Adaptive Multimodal Concept Induction}
\label{sec:induction}

Open-ended extraction can capture domain-specific and compositional semantics, but its unconstrained surface forms do not yet provide the stable identities required for catalog-wide sharing. We therefore separate concept induction into three stages: Discover expands the domain-specific semantic inventory, Standardize consolidates equivalent observations into canonical identities, and Constrain applies the resulting vocabulary consistently across the catalog.

\textbf{Discover.}
The relevant multimodal vocabulary differs across domains: movies emphasize cinematographic atmosphere and theme, while Beauty products emphasize colors, finishes, and appearance. We induce this inventory from sampled items by asking the MLLM for reusable relation--concept pairs, prioritizing coverage with candidates such as \textit{mood: whimsical}, \textit{lighting: dramatic}, or \textit{design: minimalist}. The resulting inventory is fixed for catalog-scale extraction.

\textbf{Standardize.}
Open-ended outputs contain synonymous and near-duplicate surface forms. We first group relation names using embedding-based BERTopic clustering, then use an LLM to refine clusters into canonical relations. Entity names are clustered and refined within each relation to preserve distinct semantic roles. A held-out coverage check permits a \texttt{NEW\_} option for important missing concepts. We define \emph{vocabulary mapping coverage} as the proportion of extracted relation--concept pairs assigned to the existing canonical vocabulary; unmatched pairs receive \texttt{NEW\_}. The vocabulary is updated until coverage reaches 90\%, yielding the domain vocabulary $\mathcal{E}$ and relation set $\mathcal{T}$.

The construction separates vocabulary discovery from catalog-scale extraction. Discover uses a popularity-stratified 5\% sample so that head and long-tail items both contribute candidates. Standardization and coverage validation use 20\% of the catalog; after the vocabulary converges, Constrain processes every item. All three domains reached 90\% held-out vocabulary mapping coverage within two refinement iterations. This procedure determines vocabulary size from held-out domain coverage.

\textbf{Constrain.}
Given each item's image and short title, the MLLM selects from the finalized vocabulary to produce triples $(i,r,e)$ with $i\in\mathcal{I}$, $r\in\mathcal{T}$, and $e\in\mathcal{E}$. Denote this constrained extraction by $f_{\mathrm{con}}$; then $A^{r}_{ie}=1$ exactly when $(r,e)\in f_{\mathrm{con}}(x_i)$. Items exhibiting the same concept connect to one node: for example, \texttt{mood: adventurous} connects \textit{Star Wars}, \textit{Jumanji}, and \textit{GoldenEye}, providing a common preference coordinate.

\subsection{Temporal Preference Grounding}
\label{sec:grounding}

The shared vocabulary becomes collaborative when interaction histories are mapped to the same concepts. For user $u$ and temporal bucket $t$, let $\mathcal{I}_{u,t}$ be the interacted items. We retrieve the 20 most frequent candidates within each relation, present the resulting relation-grouped pool with its evidence counts, and ask the LLM to choose one global set of at most $B=10$ concepts (exactly 10 when at least 10 candidates exist). The prompt favors concepts that are frequent and distinctive across the observed items and filters generic attributes common to nearly every candidate. The selector draws from concepts attached to the observed items, so every user--concept edge has direct support in the interaction history.

Formally, the evidence count for a candidate is
\begin{equation}
c_{u,t,r,e}=\sum_{i\in\mathcal{I}}R^{t}_{ui}A^{r}_{ie},
\qquad
\mathcal{C}_{u,t,r}=\{e:c_{u,t,r,e}>0\}.
\end{equation}
Let $g_{\theta}$ denote the constrained selection operator and $B$ its concept budget. The short-term state and its support constraint are
\begin{equation}
\begin{aligned}
P^{S,t}_{ue}&=\mathbb{I}\!\left[e\in g_{\theta}
\big(\{(r,e,c_{u,t,r,e})\}_{r,e}\big)\right],
&\|\mathbf{P}^{S,t}_{u,:}\|_0&\le B,\\
P^{S,t}_{ue}&\le\mathbb{I}\!\left[\sum_{r,i}R^{t}_{ui}A^{r}_{ie}>0\right].&&
\end{aligned}
\label{eq:grounding}
\end{equation}

We use two temporal scales. A short-term edge set $\mathcal{E}^{S}_{u,t}$ summarizes each 21-day bucket. Every $K=4$ buckets, the long-term state is recurrently updated:
\begin{equation}
\mathbf{P}^{L,t}_{u,:}=g_{\phi}\!\left(\mathbf{P}^{L,t-K}_{u,:},
\mathbf{P}^{S,t-K+1:t}_{u,:},\boldsymbol{\delta}_{u,t}\right),
\label{eq:long-grounding}
\end{equation}
where $\boldsymbol{\delta}_{u,t}$ records the age and cross-bucket frequency of each concept. The update receives the four recent short-term sets and the previous long-term set, and can output only relation--concept identities appearing in those inputs. Concepts observed in at least three of the four recent buckets are strong retention or admission candidates; concepts appearing in only one or two buckets are treated as transient, and previous concepts absent from recent evidence may be dropped. Retained concepts increase their age by 21 active days, while newly admitted concepts start at 21 days; the output contains 5--10 concepts. The graph stores $\mathbf{P}^{S}$ and $\mathbf{P}^{L}$ as typed, unweighted relations $\mathcal{R}_{S}$ and $\mathcal{R}_{L}$, allowing the recommender to learn different transformations for recent and persistent evidence.

Before a user completes four buckets, $\mathbf{P}^{L}$ is empty and the semantic encoder uses the available short-term edges. The first long-term update is initialized from those four short-term sets; later updates also receive the previous long-term state. After processing the training prefix, we export one static snapshot per user: the short-term set available after the final processed bucket and the latest long-term state. Earlier short-term sets inform recurrent long-term updates but are not unioned into the serving graph.

\textbf{A concrete semantic path.}
Suppose items in an anonymous history are attached to \textit{adventurous}, \textit{dramatic lighting}, and other canonical concepts. Retrieval exposes these supported candidates; short-term selection may retain \textit{adventurous}, while repeated occurrence across buckets may place \textit{dramatic lighting} in the long-term set. The semantic graph then contains paths such as $u\!\rightarrow\!\textit{adventurous}\!\rightarrow\!i$ for any candidate item $i$ carrying that concept. With two message-passing layers, the same concept node organizes history evidence and candidate items and directly shapes their ranking representations.

\subsection{Dual-Path Concept-Grounded Recommendation}
\label{sec:recommender}

We construct a heterogeneous graph with user, item, and concept nodes. Its relations are user--item interactions $\mathcal{R}_{\mathrm{inter}}$, item--concept edges $\mathcal{R}_{\mathrm{item}}$, and short-/long-term user--concept edges $\mathcal{R}_{S},\mathcal{R}_{L}$; reverse edges support bidirectional propagation. Separate encoders retain a dedicated collaborative representation alongside the semantic graph.

Two GAT encoders~\cite{velickovic2018graph} learn complementary paths. The CF encoder operates only on $\mathcal{R}_{\mathrm{inter}}$ and produces $\mathbf{h}_{u}^{\mathrm{CF}},\mathbf{h}_{i}^{\mathrm{CF}}$. The typed KG encoder operates on $\mathcal{R}_{\mathrm{item}}\cup\mathcal{R}_{S}\cup\mathcal{R}_{L}$ and produces $\mathbf{h}_{u}^{\mathrm{KG}},\mathbf{h}_{i}^{\mathrm{KG}}$. Before KG propagation, each concept has a scalar gate $m_e=\sigma(z_e)$ and initial representation $\widetilde{\mathbf{e}}_e=m_e\mathbf{e}_e$. Gates are initialized from graph frequency (0.3 for concepts occurring fewer than five times, 0.8 above 1,000 occurrences, and 1 otherwise) and then learned end-to-end. This gives the recommender a direct mechanism for reducing the influence of rare or overly generic concepts.

At each layer, relation-specific GAT attention normalizes over neighbors of the same edge type and then sums the transformed relation outputs. Relation-specific parameters model short-term preferences, long-term preferences, and item affiliations separately. Both encoders share learnable user/item initialization and propagate over different adjacency structures.

Separate user and item MLPs fuse the direct and semantic representations, e.g., $\mathbf{h}_u=\mathrm{MLP}_u([\mathbf{h}_{u}^{\mathrm{CF}}\|\mathbf{h}_{u}^{\mathrm{KG}}])$.
For an observed pair $(u,i^+)$ and sampled unobserved items $\mathcal{N}_u$, the ranking loss is
\begin{equation}
\mathcal{L}_{\mathrm{rec}}=-\!\sum_{(u,i^+)}\log
\frac{\exp(\mathbf{h}_u^\top\mathbf{h}_{i^+}/\tau_r)}
{\exp(\mathbf{h}_u^\top\mathbf{h}_{i^+}/\tau_r)+\sum_{j\in\mathcal{N}_u}\exp(\mathbf{h}_u^\top\mathbf{h}_{j}/\tau_r)}.
\end{equation}
Three auxiliary objectives couple the two recommendation paths and the concept structure. Let $\bar{\mathbf{h}}$ denote an $\ell_2$-normalized representation and $\mathcal{B}_c$ the user batch used for cross-view contrast. We define
\begin{equation}
s_{uv}=\bar{\mathbf{h}}_{u}^{\mathrm{CF}\top}
\bar{\mathbf{h}}_{v}^{\mathrm{KG}}/\tau_c.
\label{eq:cross-view-similarity}
\end{equation}
The resulting user-level contrastive objective is
\begin{equation}
\mathcal{L}_{\mathrm{contrast}}
=-\frac{1}{|\mathcal{B}_c|}\sum_{u\in\mathcal{B}_c}
\log\frac{\exp(s_{uu})}
{\sum_{v\in\mathcal{B}_c}\exp(s_{uv})}.
\label{eq:contrastive}
\end{equation}
Let $\mathcal{E}_{\mathrm{IC}}=\{(e,i):A_{ie}=1\}$ denote the item--concept edge set. For each $(e,i)\in\mathcal{E}_{\mathrm{IC}}$, we sample $K_{\mathrm{neg}}$ items $\mathcal{N}_{e,i}^{-}\subseteq\{j\in\mathcal{I}:A_{je}=0\}$ and define $\Delta_{eij}=\mathbf{h}_{e}^{\mathrm{KG}\top}\mathbf{h}_{i}^{\mathrm{KG}}-\mathbf{h}_{e}^{\mathrm{KG}\top}\mathbf{h}_{j}^{\mathrm{KG}}$. The alignment and gate objectives are
\begin{equation}
\mathcal{L}_{\mathrm{align}}
=-\frac{1}{K_{\mathrm{neg}}|\mathcal{E}_{\mathrm{IC}}|}
\sum_{(e,i)\in\mathcal{E}_{\mathrm{IC}}}
\sum_{j\in\mathcal{N}_{e,i}^{-}}\log\sigma(\Delta_{eij}).
\label{eq:alignment}
\end{equation}
For the concept gates, binary entropy and its vocabulary-wide mean are
\begin{equation}
\begin{aligned}
h_b(m_e)
&=-m_e\log(m_e+\epsilon)
-(1-m_e)\log(1-m_e+\epsilon),\\
H(\mathbf{m})
&=\frac{1}{|\mathcal{E}|}\sum_{e\in\mathcal{E}}h_b(m_e).
\end{aligned}
\label{eq:gate-entropy}
\end{equation}
We regularize the gates with
\begin{equation}
\mathcal{L}_{\mathrm{mask}}
=\frac{\lambda_s}{|\mathcal{E}|}\sum_{e\in\mathcal{E}}(1-m_e)
-\lambda_h H(\mathbf{m}).
\label{eq:mask}
\end{equation}
The contrastive term aligns the two history-derived user views against other users in the batch; the pairwise term aligns attached item--concept pairs against unconnected items within the semantic path. The gate objective balances concept retention and entropy. The complete objective is
\begin{equation}
\mathcal{L}=\mathcal{L}_{\mathrm{rec}}+\alpha\mathcal{L}_{\mathrm{contrast}}+
\beta\mathcal{L}_{\mathrm{align}}+\gamma\mathcal{L}_{\mathrm{mask}}.
\end{equation}
At inference, items are ranked by $\widehat{y}_{ui}=\mathbf{h}_{u}^{\top}\mathbf{h}_{i}$, keeping online computation lightweight.

\subsection{Preference Sharing through Grounded Concepts}
\label{sec:sharing-mechanism}

The item graph supplies reusable coordinates; a grounded edge $u\!\rightarrow\!e$ marks one coordinate as relevant and completes $u\!\rightarrow\!e\!\rightarrow\!i$. Eq.~\eqref{eq:semantic-match} describes the shared-concept matching route: evidence transfers when a history-grounded preference and candidate share a canonical concept. Performance should therefore depend on preserving user--concept correspondence, with the largest benefit when direct interaction evidence is limited.

\section{Experiments}
\label{sec:experiments}

We first evaluate overall recommendation quality, then test the proposed mechanism through grounding controls, activity analysis, component ablations, vocabulary transfer, and construction robustness.

\begin{table*}[t]
\centering
\small
\setlength{\tabcolsep}{4.2pt}
\caption{Recommendation performance, reported as mean $\pm$ standard deviation over five downstream training seeds. ML-1M uses full ranking; Video Games and Beauty use uni100. Bold and underline denote the best and second-best mean. $^*$ marks a significant improvement over the strongest baseline reported in the corresponding experiment (paired $t$-test, $p<0.05$).}
\label{tab:main}
\resizebox{\textwidth}{!}{%
\begin{tabular}{l|cc|cc|cc}
\toprule
\multirow{2}{*}{Method} & \multicolumn{2}{c|}{ML-1M (full)} & \multicolumn{2}{c|}{Video Games (uni100)} & \multicolumn{2}{c}{Beauty (uni100)} \\
& NDCG@10 & Recall@10 & NDCG@10 & Recall@10 & NDCG@10 & Recall@10 \\
\midrule
BPR      & 0.1209$\pm$.0021 & 0.0595$\pm$.0007 & 0.3453$\pm$.0013 & 0.5195$\pm$.0018 & 0.5089$\pm$.0041 & 0.5928$\pm$.0072 \\
LightGCN & 0.1262$\pm$.0014 & 0.0589$\pm$.0007 & 0.3567$\pm$.0014 & 0.5574$\pm$.0027 & 0.5311$\pm$.0098 & 0.6106$\pm$.0154 \\
KGAT     & 0.1195$\pm$.0013 & 0.0594$\pm$.0007 & 0.3532$\pm$.0016 & 0.5397$\pm$.0027 & 0.5185$\pm$.0087 & 0.5998$\pm$.0149 \\
\midrule
VBPR     & 0.1095$\pm$.0011 & 0.0459$\pm$.0016 & 0.3551$\pm$.0072 & 0.5378$\pm$.0099 & 0.5267$\pm$.0073 & 0.6039$\pm$.0029 \\
MMGCN    & \underline{0.1329$\pm$.0012} & 0.0556$\pm$.0004 & \underline{0.3569$\pm$.0032} & 0.5690$\pm$.0025 & 0.5592$\pm$.0079 & 0.6609$\pm$.0081 \\
MKGAT    & 0.1317$\pm$.0009 & 0.0554$\pm$.0006 & 0.3354$\pm$.0017 & 0.5335$\pm$.0035 & 0.5574$\pm$.0105 & 0.6687$\pm$.0092 \\
LATTICE  & 0.1037$\pm$.0002 & 0.0528$\pm$.0004 & 0.3507$\pm$.0024 & 0.5655$\pm$.0035 & 0.5281$\pm$.0108 & 0.6484$\pm$.0156 \\
BM3      & 0.1073$\pm$.0019 & 0.0440$\pm$.0006 & 0.3559$\pm$.0003 & 0.5623$\pm$.0005 & 0.5344$\pm$.0122 & 0.6572$\pm$.0143 \\
FREEDOM  & 0.1175$\pm$.0011 & 0.0563$\pm$.0003 & 0.3464$\pm$.0005 & 0.5616$\pm$.0014 & 0.5156$\pm$.0008 & 0.6376$\pm$.0013 \\
MONET    & 0.1126$\pm$.0015 & 0.0550$\pm$.0007 & 0.3506$\pm$.0011 & 0.5495$\pm$.0013 & 0.5434$\pm$.0107 & 0.6628$\pm$.0146 \\
SMORE    & 0.1244$\pm$.0014 & 0.0584$\pm$.0009 & 0.3519$\pm$.0021 & 0.5593$\pm$.0034 & 0.5511$\pm$.0096 & 0.6648$\pm$.0138 \\
\midrule
TALLRec  & 0.1271$\pm$.0075 & \underline{0.0602$\pm$.0013} & 0.3201$\pm$.0057 & 0.5620$\pm$.0086 & 0.5483$\pm$.0059 & 0.6802$\pm$.0061 \\
DOGE     & 0.1082$\pm$.0015 & 0.0535$\pm$.0006 & 0.3186$\pm$.0051 & 0.5175$\pm$.0036 & 0.5484$\pm$.0154 & \underline{0.6884$\pm$.0129} \\
MLLMRec  & 0.1132$\pm$.0033 & 0.0587$\pm$.0011 & 0.3526$\pm$.0042 & 0.5666$\pm$.0073 & 0.5595$\pm$.0081 & 0.6759$\pm$.0079 \\
MLLM-MSR & 0.1200$\pm$.0057 & 0.0510$\pm$.0020 & 0.3509$\pm$.0049 & \underline{0.5722$\pm$.0104} & \underline{0.5638$\pm$.0156} & 0.6575$\pm$.0066 \\
\midrule
\textbf{MKGE} & \textbf{0.1336$\pm$.0022} & \textbf{0.0619$^*\!\pm$.0013} & \textbf{0.3619$^*\!\pm$.0069} & \textbf{0.5766$^*\!\pm$.0074} & \textbf{0.5789$^*\!\pm$.0127} & \textbf{0.7219$^*\!\pm$.0121} \\
\bottomrule
\end{tabular}}
\vspace{5pt}
\end{table*}

\subsection{Experimental Setup}

\textbf{Datasets and splits.}
We evaluate on MovieLens-1M (ML-1M)~\cite{10.1145/2827872}, Amazon Video Games, and Amazon Beauty~\cite{ni-etal-2019-justifying}. ML-1M and Video Games use 5-core filtering; Beauty retains the standard unfiltered subset. Table~\ref{tab:data-stats} summarizes the resulting scale. Movie posters follow the public multimodal resource used by MMGCN~\cite{10.1145/3343031.3351034}; Amazon product images and titles come from public metadata. We order each user's interactions chronologically and apply a per-user 70/10/20 split using integer cut points, preserving a nonempty training prefix and test suffix for histories with at least three interactions; histories shorter than three remain entirely in training. At the smallest valid history length the validation slice may be empty. Item--concept induction uses catalog content, while user grounding uses only the training prefix.

\begin{table}[t]
\centering
\small
\setlength{\tabcolsep}{4pt}
\caption{Dataset statistics after preprocessing.}
\label{tab:data-stats}
\resizebox{\columnwidth}{!}{%
\begin{tabular}{lrrrrrr}
\toprule
Dataset & Users & Test users & Items & Interactions & Density & Int./user \\
\midrule
ML-1M & 6,040 & 6,040 & 3,416 & 999,611 & 4.84\% & 165.5 \\
Video Games & 55,211 & 54,692 & 14,969 & 446,532 & 0.054\% & 8.09 \\
Beauty & 233,222 & 3,192 & 16,340 & 264,801 & 0.007\% & 1.14 \\
\bottomrule
\end{tabular}}
\end{table}

The datasets span sharply different interaction regimes, letting us test whether shared concept paths become more useful as direct collaborative evidence weakens. Beauty's 3,192 test users remain sparse after conditioning on a valid test suffix: their training histories contain 2.15 interactions on average (median 2; interquartile range 2--2). These evaluable users account for 1.37\% of all users in the preprocessed dataset.

\begingroup\sloppy
\textbf{Baselines.}
We compare with four groups of recommenders. BPR~\cite{rendle2009bpr} and LightGCN~\cite{10.1145/3397271.3401063} provide collaborative anchors, while KGAT~\cite{10.1145/3292500.3330989} measures knowledge-aware propagation. VBPR, MMGCN, MKGAT, LATTICE, BM3, FREEDOM, MONET, and SMORE cover continuous multimodal matching, graph structure learning, target-aware user modeling, self-supervision, and multimodal fusion. TALLRec provides a tuned language-model anchor; DOGE combines LLM-enhanced multimodal features with hypergraph propagation; and MLLM-MSR and MLLMRec summarize or reason over multimodal user preferences. We use RecBole for BPR, LightGCN, and KGAT and official implementations for the rest, tuning shared hyperparameters on validation data.
\par\endgroup

\textbf{Protocols.}
We report NDCG@10 and Recall@10 over five random seeds with paired $t$-tests. ML-1M uses full ranking. For each held-out Amazon positive, uni100 ranks it against 99 negatives sampled uniformly without replacement, excluding all of the user's training, validation, and test positives. Metrics are averaged over these positive-specific queries. Candidate sets are shared across methods within each seed, including large-model baselines with per-candidate user-conditioned inference.

Figure~\ref{fig:grounding-controls} reports a separate three-seed grounding diagnostic on ML-1M and Video Games under a common BF16 profile, batch size 8,192, matched sampled candidates, and uni100 evaluation. Figure~\ref{fig:case-study} compares Full and No User KG over the full ML-1M catalog. In the exported static graphs, nonempty short-/long-term states cover 99.5\%/99.9\% of ML-1M test users and 90.1\%/90.4\% of Video Games test users, averaging 9.92/5.00 and 4.66/3.33 edges per test user, respectively.

\textbf{Implementation.}
The recommender uses 64-dimensional embeddings, two GNN layers, four attention heads, dropout 0.2, batch size 2,048, and one sampled negative per positive interaction. Adam optimizes the model with learning rate $10^{-3}$ and weight decay $10^{-5}$ for at most 300 epochs; validation NDCG@10 controls early stopping with patience 10. The ranking and contrast temperatures are 0.2 and 0.1, $(\alpha,\beta,\gamma)=(0.1,0.05,0.01)$, $(\lambda_s,\lambda_h)=(1.0,0.1)$, and $K_{\mathrm{neg}}=5$. GPT-4o-mini uses temperature 0 for the main construction. Standardization uses all-MiniLM-L6-v2 embeddings and BERTopic with minimum topic size 10.

\subsection{Main Results}

Table~\ref{tab:main} shows that MKGE achieves the best result on all six metrics under the respective protocols. On full-ranking ML-1M, it improves over the strongest baseline by 0.53\% in NDCG and 2.82\% in Recall. The improvements are 1.40\%/0.77\% on Video Games and 2.68\%/4.87\% on Beauty for NDCG/Recall. 

\begingroup\sloppy
Against the stronger result of MONET and SMORE on each dataset, MKGE improves NDCG@10/Recall@10 by 7.40\%/5.99\% on ML-1M, 2.84\%/3.09\% on Video Games, and 5.04\%/8.59\% on Beauty. It also outperforms DOGE, MLLMRec, and MLLM-MSR. These gains extend across representation-based and large-model baselines; the following controls examine the role of shared factor identities and personalized grounding.
\par\endgroup

\subsection{Does Grounding Encode the Right Preferences?}

We construct four controls with the same item vocabulary and recommender. \textit{No User KG} removes every user--concept edge. \textit{Frequency} selects the most frequent concepts in each observed history. \textit{IDF} ranks a history-exposed concept $e$ by $\operatorname{freq}(u,e)\log[(|\mathcal{U}|+1)/(n_e+1)]$, where $n_e$ is the number of training users exposed to $e$. Both heuristics match Full's per-user short-/long-term edge budget. \textit{Rewired} independently randomizes the two bipartite graphs, exactly preserving user and concept degrees within each edge type and disrupting their original correspondence.

\begin{figure}[t]
\centering
\includegraphics[width=\columnwidth]{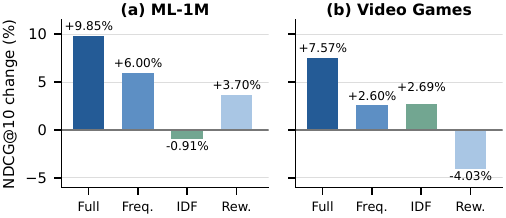}
\vspace{-4pt}
\caption{Relative NDCG@10 change over No User KG. Freq., IDF, and Rew. denote frequency, inverse-document-frequency, and degree-preserving rewiring; Full is strongest on both datasets.}
\Description{Two side-by-side bar charts show relative NDCG at 10 changes over No User KG for Full grounding, Frequency, IDF, and Rewired on ML-1M and Video Games.}
\label{fig:grounding-controls}
\end{figure}

In Fig.~\ref{fig:grounding-controls}, Full grounding exceeds Frequency by 3.63\% on ML-1M and 4.84\% on Video Games, IDF by 10.86\% and 4.75\%, Rewired by 5.93\% and 12.09\%, and No User KG by 9.85\% and 7.57\%. IDF is effectively tied with Frequency on Video Games (.3385 versus .3382) and falls below it on ML-1M (.2476 versus .2649). All four ML-1M comparisons and the Video Games comparison with Rewired are significant under paired $t$-tests ($p<.05$).

\begin{figure}[t]
\centering
\includegraphics[width=.82\columnwidth]{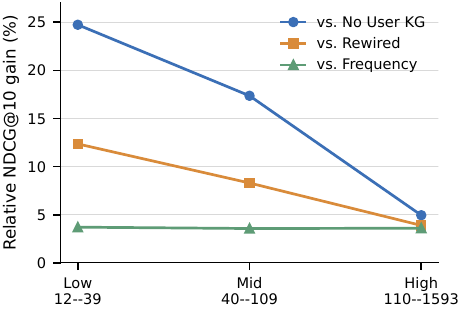}
\vspace{-4pt}
\caption{Relative NDCG@10 improvement of full grounding over each control by ML-1M activity tier. The benefit over removing or rewiring the user graph is largest for low-activity users.}
\Description{Three lines show the NDCG gain of full grounding over No User KG, Rewired, and Frequency for low-, mid-, and high-activity users. Every line decreases with activity.}
\label{fig:activity}
\end{figure}

\textbf{Grounding helps most when individual evidence is limited.}
Fig.~\ref{fig:activity} groups ML-1M users into roughly equal activity tiers and reports relative improvements. Full grounding improves over No User KG by 24.74\%, 17.36\%, and 4.95\% from low to high activity. Its relative advantages over Rewired are 12.36\%, 8.30\%, and 3.89\%. The improvement over Frequency remains approximately 3.6\% across tiers.

\begin{table*}[t]
\centering
\small
\caption{Component ablations under uni100. Each cell reports NDCG@10 / Recall@10. $\dagger$ marks a nonsignificant degradation; all other degradations are significant ($p<.05$).}
\label{tab:component-ablation}
\begin{minipage}[t]{.485\textwidth}
\centering
\textbf{(a) Recommendation architecture}\par\smallskip
\resizebox{\linewidth}{!}{%
\begin{tabular}{lccc}
\toprule
Variant & ML-1M & Video Games & Beauty \\
\midrule
Full MKGE & \textbf{.2707/.5006} & \textbf{.3619/.5766} & \textbf{.5789/.7219} \\
CF path only & .2652/.4931 & .3403/.5543 & .5219/.6843 \\
Semantic path only & .2410/.4502 & .3337/.4975 & .5583/.6996 \\
w/o contrast & .2639/.4894 & .3186/.5281 & .5210/.6944 \\
w/o entity mask & .2690$^\dagger$/.4973$^\dagger$ & .3316/.5411 & .5303/.6910 \\
\bottomrule
\end{tabular}}
\end{minipage}\hfill
\begin{minipage}[t]{.485\textwidth}
\centering
\textbf{(b) Grounded knowledge}\par\smallskip
\resizebox{\linewidth}{!}{%
\begin{tabular}{lccc}
\toprule
Variant & ML-1M & Video Games & Beauty \\
\midrule
Full MKGE & \textbf{.2707/.5006} & \textbf{.3619/.5766} & \textbf{.5789/.7219} \\
w/o user edges & .2447/.4605 & .3491/.5572 & .4332/.6791 \\
w/o item edges & .2699$^\dagger$/.5002 & .3531/.5497 & .4909/.6831 \\
w/o short-term & .2650/.4895 & .3486/.5459 & .5352/.6941$^\dagger$ \\
w/o long-term & .2596/.4786 & .3462/.5508 & .5241/.6932 \\
\bottomrule
\end{tabular}}
\end{minipage}
\end{table*}

\textbf{Both recommendation paths matter.}
Table~\ref{tab:component-ablation}(a) separates direct and concept-mediated recommendation. Removing either path lowers performance across datasets. Their balance changes with sparsity: on dense ML-1M the direct path is stronger, while on Beauty the semantic path retains more of the full model's performance. Cross-path contrast produces larger drops on the Amazon datasets.

\textbf{User grounding drives personalization.}
Table~\ref{tab:component-ablation}(b) shows that removing user--concept edges is more damaging than removing item--concept edges on every dataset, including a 25.2\% NDCG drop on Beauty. Item--concept edges contribute increasingly on the sparser datasets, reaching a 15.2\% NDCG contribution on Beauty.

The two horizons organize preference by persistence rather than partitioning concepts into unrelated sets. On ML-1M, each user has 9.9 short-term and 5.0 long-term concepts on average; 98.8\% of long-term links also appear in the final short-term profile, and 99.5\% of users share at least one concept across horizons. Long-term grounding therefore retains a more selective recurrent core, while the short-term state preserves a broader view of the latest activity. Removing either horizon degrades all three datasets, with larger NDCG losses from removing the long-term state, supporting distinct graph relations for recent and persistent evidence.

Each knowledge ablation removes the named edge set and retains the shared vocabulary and remaining graph relations, measuring its marginal contribution in a common semantic coordinate system.

\textbf{Canonical identity is useful.}
Bypassing Standardize and using open-ended MLLM outputs directly lowers Video Games NDCG@10 from $.3619\pm.0069$ to $.3510\pm.0038$ and Recall@10 from $.5766\pm.0074$ to $.5662\pm.0042$. Consolidating recurring surface forms strengthens the paths that reuse them.

\subsection{Case Study: From Grounded Evidence to Ranking}

\begin{figure*}[t]
  \centering
  \includegraphics[width=.94\textwidth]{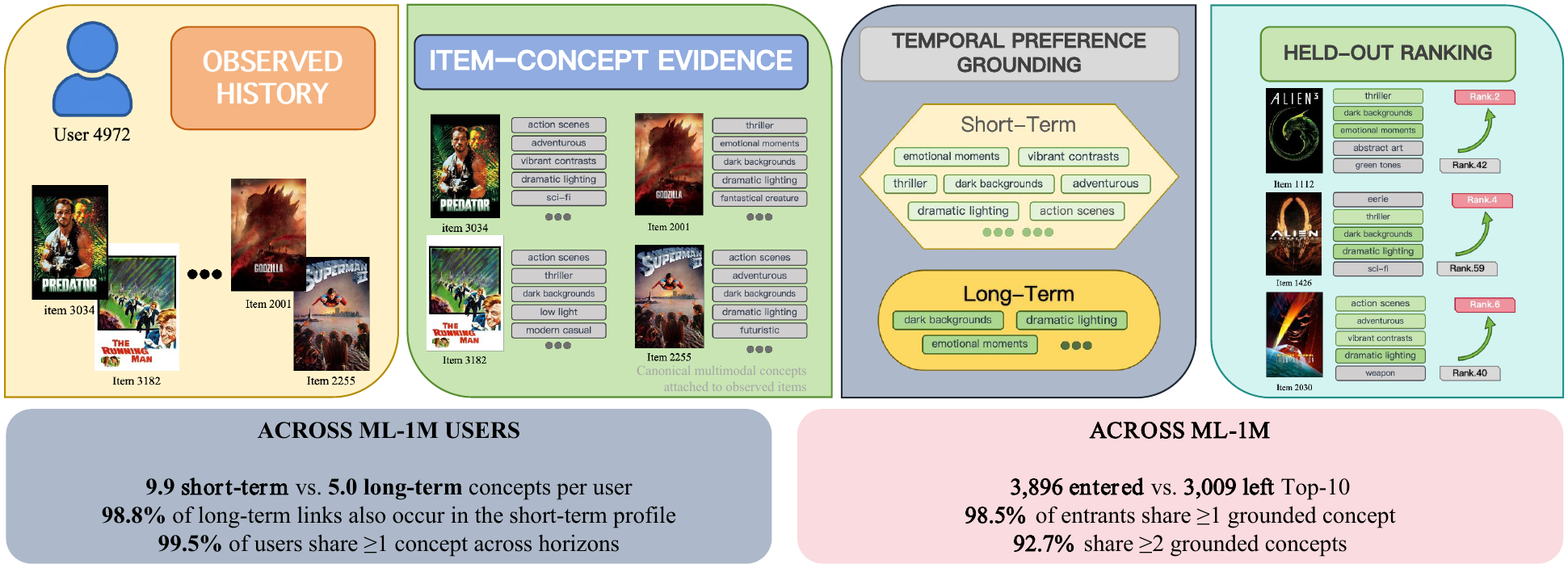}
  \vspace{-4pt}
  \caption{ML-1M case study. Observed item--concept evidence grounds user 4972's short- and long-term preferences, promoting three held-out positives into the full-catalog Top-10. Bottom insets summarize cross-horizon concept overlap and paired Full/No User KG transitions over the dataset.}
  \Description{The upper row traces four observed history items through their multimodal concepts and the user's short- and long-term grounded preferences to three held-out items whose full-catalog ranks improve from 42, 59, and 40 to 2, 4, and 6. The lower row summarizes temporal-profile overlap and Top-10 entries and exits across ML-1M.}
  \label{fig:case-study}
\end{figure*}

Fig.~\ref{fig:case-study} combines population context with one anonymous user trace from the paired Full/No User KG runs. Across all 205,337 held-out positives, 3,896 items enter the full-catalog Top-10 under Full while 3,009 leave it, giving a net increase of 887. Of the entering items, 3,839 share at least one grounded concept with their user and 3,612 share at least two. For user 4972, repeated history evidence supports a coherent action/thriller profile at short and long horizons. Three representative held-out positives sharing these concepts move from ranks 40--59 under No User KG to ranks 2--6 under Full. These paired rank changes trace how history-supported concepts connect the user to held-out items under Full.

\begin{figure}[t]
  \centering
  \includegraphics[width=\columnwidth]{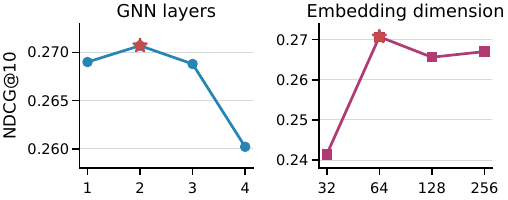}
  \vspace{-4pt}
  \caption{Sensitivity on ML-1M: (left) GNN depth and (right) embedding dimension. Two layers and 64 dimensions perform best among the tested settings.}
  \Description{Two compact side-by-side line charts show NDCG at 10 for one to four GNN layers and embedding dimensions 32, 64, 128, and 256.}
  \label{fig:hyperparameters}
  \vspace{5pt}
\end{figure}

\textbf{Performance is stable over moderate model capacities.}
Fig.~\ref{fig:hyperparameters} reports the architecture sensitivity analysis. One layer is insufficient for the two-hop user--concept--item route, and four layers perform worse. Using 32 dimensions reduces performance; results remain stable from 64 to 256 dimensions. Temporal windows of 14, 21, and 30 days yield NDCG@10 values of .2704, .2707, and .2700.

\subsection{Why Is a Shared Vocabulary Needed?}

The induced vocabulary connects multiple items and user histories to each canonical concept. Table~\ref{tab:graph-scale} summarizes the resulting scale and reuse.

\begin{table}[t]
\centering
\small
\setlength{\tabcolsep}{4.0pt}
\caption{Scale and reuse of the induced shared vocabulary. I--C and U--C denote item--concept and user--concept edges.}
\label{tab:graph-scale}
\resizebox{\columnwidth}{!}{%
\begin{tabular}{lrrrr}
\toprule
Dataset & Concepts & I--C edges & I--C/concept & U--C edges \\
\midrule
ML-1M & 155 & 32,675 & 210.8 & 90,100 \\
Video Games & 139 & 61,648 & 443.5 & 437,932 \\
Beauty & 453 & 88,379 & 195.1 & 167,035 \\
\bottomrule
\end{tabular}}
\end{table}

Each concept participates in hundreds of item edges on average, allowing one grounded identity to connect a user history to otherwise different items. On ML-1M, 93 concepts connect at least 50 items and 76 connect at least 100. \textit{Bright lighting}, \textit{dramatic lighting}, and \textit{adventurous} connect 1,068, 974, and 554 items, respectively. The larger user--concept graphs support personalized selections from this shared vocabulary. Induction remains domain-adaptive: movies emphasize atmosphere and cinematography, games emphasize characters and game aesthetics, and Beauty emphasizes concrete product appearance.

\begin{figure}[t]
  \centering
  \includegraphics[width=.78\columnwidth]{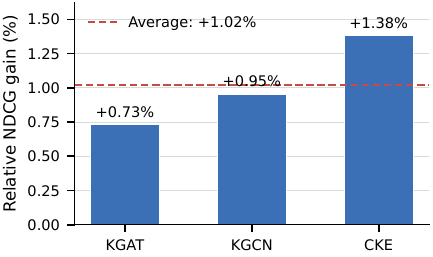}
  \vspace{-4pt}
  \caption{Replacing metadata with MKGE item--concept triples improves three KG recommenders on ML-1M without architectural changes.}
  \Description{A bar chart shows relative NDCG gains of 0.73, 0.95, and 1.38 percent for KGAT, KGCN, and CKE, with an average of 1.02 percent.}
  \label{fig:transfer}
\end{figure}

Fig.~\ref{fig:transfer} isolates the item graph from MKGE's own recommender and user grounding. The original ML-1M metadata graph contains \textit{Genre} and \textit{Year} relations; replacing those triples with MKGE item--concept triples improves KGAT, KGCN~\cite{wang2019kgcn}, and CKE~\cite{zhang2016cke} without changing their architectures, for a 1.02\% average NDCG gain. The induced multimodal concepts therefore remain useful beyond MKGE's dual-path recommender.

\subsection{Robustness, Efficiency, and Reproducibility}

In Table~\ref{tab:construction-model}, Qwen3-VL-8B replaces GPT-4o-mini throughout concept induction and preference grounding, constructing a separate graph. Its NDCG@10 differs by 0.67\%, 0.47\%, and 1.00\% on ML-1M, Video Games, and Beauty. The shared-vocabulary interface thus preserves similar downstream quality across construction models: the recommender consumes normalized concepts and graph relations rather than model-specific generation traces. The released graphs use the original GPT-based pipeline.

\begin{table}[H]
\centering
\small
\setlength{\tabcolsep}{5pt}
\caption{Robustness to the construction model (NDCG@10).}
\label{tab:construction-model}
\begin{tabular}{lccc}
\toprule
Construction model & ML-1M & Video Games & Beauty \\
\midrule
GPT-4o-mini & .1336 & .3619 & .5789 \\
Qwen3-VL-8B & .1327 & .3602 & .5731 \\
\bottomrule
\end{tabular}
\end{table}

Construction and serving costs are separated by design. Item extraction takes 4.9 seconds per item and preference grounding 2.2 seconds per user; with 20 parallel workers, initial ML-1M construction takes approximately 25 minutes. Across the three datasets, the recorded API cost is below USD 100. Online recommendation then requires only the GNN forward pass, with the comparison summarized in Table~\ref{tab:cost}.

\begin{table}[H]
\centering
\small
\setlength{\tabcolsep}{3.5pt}
\caption{Training and online inference cost on ML-1M. Offline MKGE construction is reported separately in text.}
\label{tab:cost}
\begin{tabular}{lrrr}
\toprule
Method & Train/epoch & ms/user & GPU memory \\
\midrule
TALLRec & 20 min & 5,210 & 14.9 GB \\
MLLM-MSR & 23 min & 3,290 & 22.9 GB \\
MKGE recommender & 52 s & 51.3 & 8.2 GB \\
\bottomrule
\end{tabular}
\end{table}

The same separation supports incremental maintenance. New items use Constrain against the current vocabulary, while new interactions update the current user bucket; persistent vocabulary growth and four completed buckets trigger rediscovery and long-term grounding, respectively. Each update touches the affected items, users, or scheduled vocabulary state.

We release the constructed KGs for all three datasets, the exact extraction and grounding prompts, and the downstream recommendation implementation at \url{https://github.com/Garden1a-X/MKGE}. The released graph files allow direct reproduction of the recommender experiments from fixed construction outputs.

\FloatBarrier

\section{Discussion and Conclusion}

MKGE connects behavioral evidence with candidate items through shared canonical concepts. Grounding outperforms frequency, IDF, and topology-preserving controls, with the largest ML-1M gains for low-activity users. The induced item graph also improves three established KG recommenders, supporting its utility across architectures.

Grounded concepts summarize history-supported preference associations with observable multimodal factors. Direct collaborative representations complement this view with signals related to price, popularity, social influence, and latent behavioral structure. Domain changes may require vocabulary refresh, and concept quality remains linked to multimodal recognition and normalization. Broader domains and evolving catalogs offer natural directions for future evaluation. Explicit concept profiles also motivate studying user inspection and revision. Further extensions could examine how shared semantic factors support preference-guidance paths~\cite{hou2026prorl} and user recommendation that accounts for information propagation~\cite{hou2025heterogeneous}. MKGE shows how multimodal item understanding can become a shared language for user preference modeling.

More broadly, stable factor identities let catalog construction, temporal grounding, and recommendation evolve separately through a shared semantic interface. This offers a route for turning increasingly capable multimodal understanding into structured user models that remain reusable across candidates and downstream architectures.\label{end:main-content}

\section{Ethical Considerations}
This work uses public benchmark datasets. MKGE summarizes historical interactions as user--concept edges, which may reflect recognition errors and representational biases inherited from the construction model and source media. Responsible deployment requires compliance with data licenses, concept audits for target populations, and mechanisms for users to inspect, update, or delete stored preference representations.

\clearpage
\bibliographystyle{ACM-Reference-Format}
\bibliography{references}

\end{document}

%% file: author_info.tex
\author{Ao Xu}
\orcid{0009-0005-9080-4882}
\affiliation{%
  \department{School of Data Science}
  \institution{Fudan University}
  \city{Shanghai}
  \country{China}
}
\email{axu24@m.fudan.edu.cn}
\author{Jinhui Huang}
\orcid{0009-0007-6216-8684}
\affiliation{%
  \department{School of Data Science}
  \institution{Fudan University}
  \city{Shanghai}
  \country{China}
}
\email{jhhuang25@m.fudan.edu.cn}
\author{Wenqiang Tang}
\orcid{0009-0004-6787-9934}
\affiliation{%
  \department{School of Data Science}
  \institution{Fudan University}
  \city{Shanghai}
  \country{China}
}
\email{tangwq26@m.fudan.edu.cn}
\author{Tiehua Mei}
\orcid{0009-0005-9677-4653}
\affiliation{%
  \department{School of Data Science}
  \institution{Fudan University}
  \city{Shanghai}
  \country{China}
}
\email{thmei24@m.fudan.edu.cn}
\author{Hongru Hou}
\orcid{0000-0003-3471-3781}
\affiliation{%
  \department{School of Data Science}
  \institution{Fudan University}
  \city{Shanghai}
  \country{China}
}
\email{hrhou25@m.fudan.edu.cn}
\author{Hengrui Chen}
\orcid{0009-0002-3310-9224}
\affiliation{%
  \department{School of Data Science}
  \institution{Fudan University}
  \city{Shanghai}
  \country{China}
}
\email{chenhr24@m.fudan.edu.cn}
\author{Jiaqing Liang}
\orcid{0000-0003-0670-5602}
\affiliation{%
  \department{School of Data Science}
  \institution{Fudan University}
  \city{Shanghai}
  \country{China}
}
\email{liangjiaqing@fudan.edu.cn}
\author{Deqing Yang}
\authornote{Corresponding author.}
\orcid{0000-0002-1390-3861}
\affiliation{%
  \department{School of Data Science}
  \institution{Fudan University}
  \city{Shanghai}
  \country{China}
}
\email{yangdeqing@fudan.edu.cn}